\documentclass[a4paper]{jpconf}
\usepackage{graphicx}
\usepackage{listings}
\usepackage{color}
\usepackage{subfig}

\begin{document}
\title{An open-source job management framework for parameter-space exploration: OACIS}

\author{Y. Murase$^1$, T. Uchitane$^2$ and N. Ito$^{1,3}$}

\address{$^1$RIKEN Advanced Institute for Computational Science, Kobe, Hyogo 650-0047, Japan}
\address{$^2$Research Institute for Economics and Business Administration, Kobe University, Kobe, Hyogo 657-8501, Japan}
\address{$^3$Department of Applied Physics, Graduate School of Engineering, The University of Tokyo, Bunkyo-ku, Tokyo 113-8656, Japan}

\ead{$^1$yohsuke.murase@gmail.com, $^2$uchitane@rieb.kobe.ac.jp}

\begin{abstract}
We present an open-source software framework for parameter-space exporation, named OACIS, which is useful to manage vast amount of simulation jobs and results in a systematic way.
Recent development of high-performance computers enabled us to explore parameter spaces comprehensively, however, in such cases, manual management of the workflow is practically impossible.
OACIS is developed aiming at reducing the cost of these repetitive tasks when conducting simulations by automating job submissions and data management.
In this article, an overview of OACIS as well as a getting started guide are presented.
\end{abstract}

\section{Introduction}

In this paper we present OACIS (Organizing Assistant for Comprehensive and Interactive Simulations) \cite{Murase2014Tool}, an open-source job management software for large-scale simulations (\verb"http://github.com/crest-cassia/oacis").
When we conduct numerical simulations in scientific research, we usually carry out many simulation jobs changing models and parameters by trial and error.
This kind of trial-and-error approach often causes a problem of job management since a large amount of jobs are often created while we are exploring the parameter space.
For instance, submitting one simulation job would require the following tasks.

\begin{itemize}
  \item login to a remote host
  \item make a directory for a job
  \item create a shell script to submit a job
  \item submit your job to the job-scheduler
  \item wait until the job finishes
  \item transfer the result files to the local machine
  \item conduct statistical analysis on the results
  \item make plots
\end{itemize}

This process is repeated until he or she gets satisfactory results.
Such repetitions are not only troublesome and tedious but prone to human errors.
In order to conduct research in an efficient, reliable, and reproducible way, a software to help management of simulation results are strongly expected.

In case of social simulation, the problem can be even more critical because the amount of parameter space exploration is often larger than those for physical or chemical phenomena \cite{noda2015roadmap}.
One of the significant difficulty in social simulation is the lack of established model.
The models for physical simulation are usually based on solid first principle equations, such as Schr\"{o}dinger equation or Navier-Stokes equation, which are carefully verified by real experiments.
On the other hand, the models for social simulation are inherently phenomenological ones.
These models are often constructed based on the observation of real society and/or our common senses.
Even though an increasing amount of data about society is available, these data are inherently incomplete due to technical and privacy reasons.
Therefore social simulations are inevitably less reliable compared to conventional physical simulations hence a single simulation run is just one of the possible futures.
To obtain robust and universal information which is independent of minute details of model assumptions, we need to investigate various models and understand the global landscape of these models, which requires us to explore vast parameter spaces.

Using OACIS, we can automate most of the above tasks, which lets you efficiently explore the parameter space.
With a user-friendly interface, you can easily submit various jobs to appropriate remote hosts.
After these jobs are finished, all the result files are automatically downloaded from the remote hosts and stored in a traceable way together with logs of the date, host, and elapsed time of the jobs.
You can easily find the status of the jobs or results files from a web browser, which lets you focus on more productive and essential parts of your research activities.

The development of OACIS has started in April of 2013 and is released as an open-source software in 2014 under the MIT license.
After the first release, minor version-upgrades have been iteratively released which includes various improvements and bug fixes.
In Oct.~2015, a major version upgrade, version 2.0, was released which includes significant enhancements.
It has been applied to various research projects including studies on network science \cite{Murase2014Multilayer,Murase2015Modeling,torok2016big}, molecular dynamics simulations\cite{Kuwabara2016JPhys}, and agent simulations \cite{torii2015shock} to name a few.
Functions needed by these projects have been implemented and usability are validated in actual projects.

\section{Overview}

\subsection{Data Structure}

\begin{figure}[ht]
\begin{center}
\includegraphics[width=0.75\columnwidth]{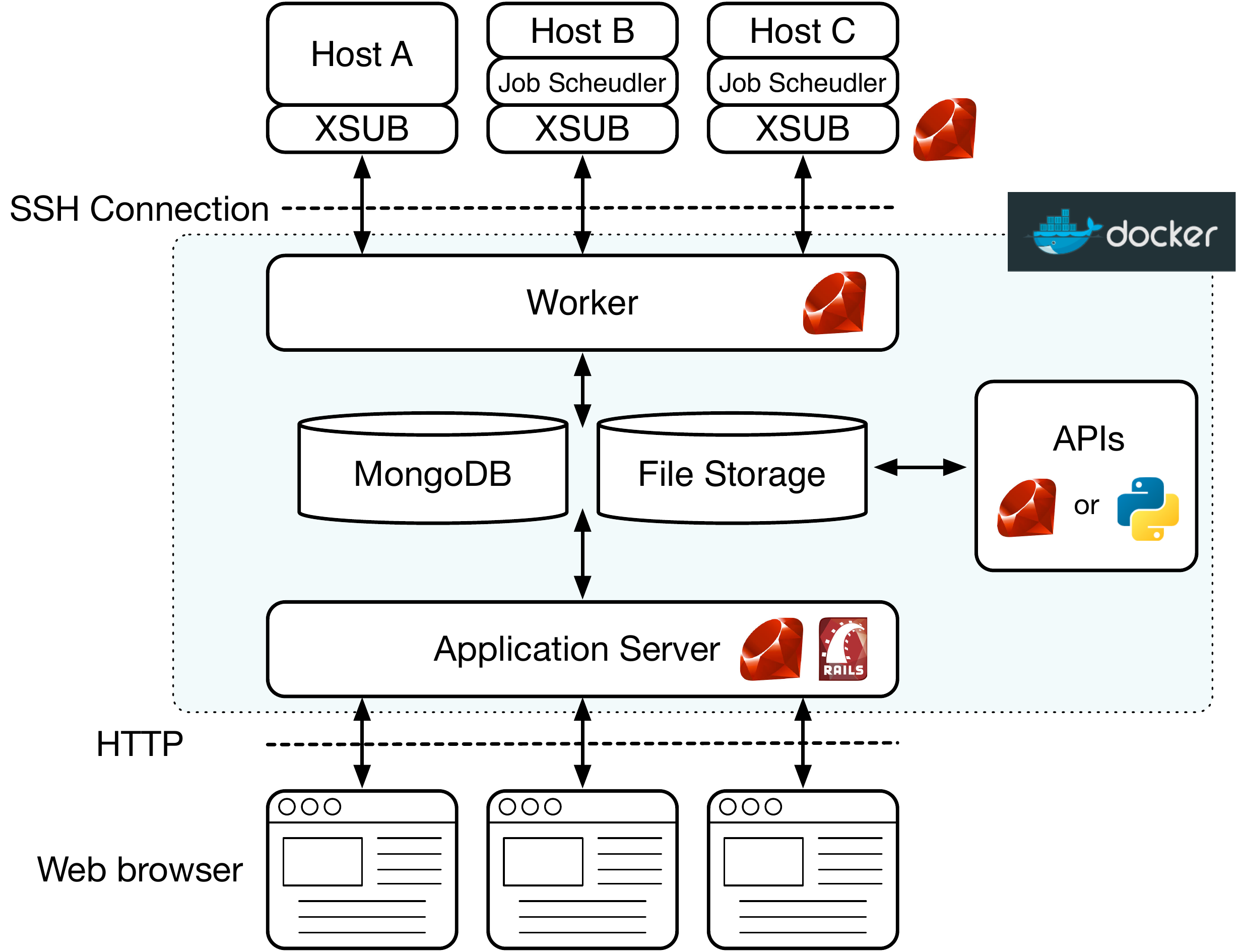}
\caption{\label{fig:architecture}
An overview of the system architecture of OACIS. The application server developed on Ruby on Rails framework provides a user interface.
The request from users are stored in the MongoDB and they are handled by another daemon process called a ``worker''.
The worker process makes an SSH connection to remote hosts and submits jobs.
After the jobs finished, worker downloads results and stored them into the database and the file system.
Users can also use Python or Ruby scripts to operate OACIS.
The Ruby logo by Yukihiro Matsumoto is licensed under CC BY-SA 2.5.
}
\end{center}
\end{figure}

The system architecture of OACIS is depicted in Fig.~\ref{fig:architecture}.
The web server is developed based on the Ruby on Rails framework (\verb"http://rubyonrails.org/"), which provides an interactive user-interface.
MongoDB (\verb"http://www.mongodb.org/"), a document-based database, is used as data storage.
The application server is responsible for handling requests from users.
When a user creates a job using a web browser, the record of the job is created in the database.
Another daemon process, which we call a ``worker'', periodically checks whether a job is ready to be submitted to a remote host.
If a job is found, the worker generates a shell script to execute a job and submits it to the job scheduler on the remote host via SSH connection.
Hereafter, we call these remote hosts ``computational hosts''.
Then the worker process periodically checks the status of the submitted jobs.
When one of the submitted jobs is finished, the worker downloads the results and stores them into the file storage and the database in an organized way.
Hence, users do not have to check the status of the remote hosts by themselves and they can trace the simulation results even after several months.
Various logs about the jobs, including the values of parameters, executed dates, elapsed times, are automatically kept by OACIS and they are stored in the database.
On the other hand, the files generated from the simulators are stored on a file system, which enables you to directly access the simulation outputs using a file browser or a command line terminal.

\begin{figure}[ht]
\begin{center}
\includegraphics[width=0.7\columnwidth]{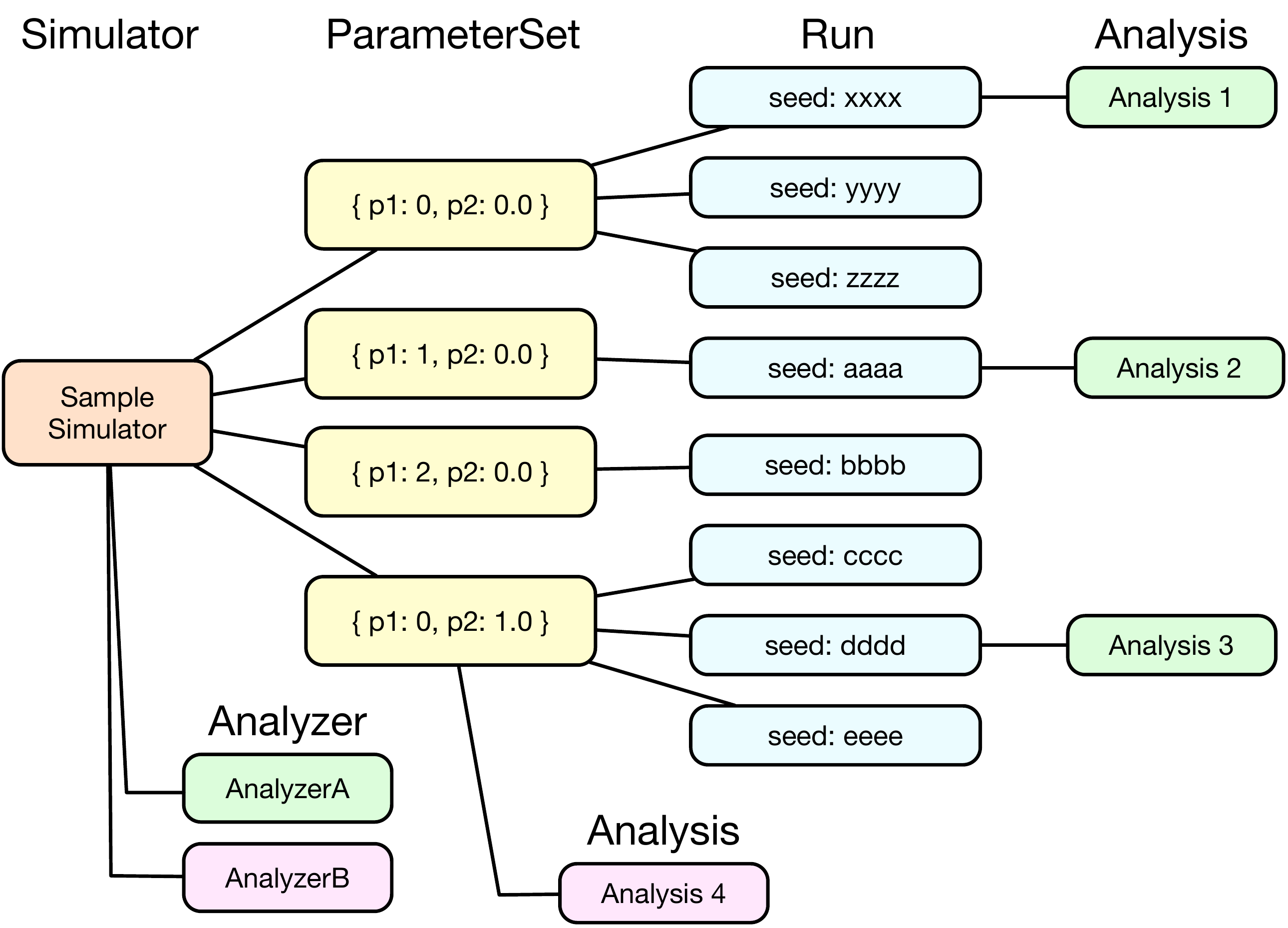}
\caption{\label{fig:data-structure}
A hierarchical data structure is used by OACIS. On top layer, ``Simulator'' objects exists which corresponds to an executable program.
``ParameterSet'' corresponds to a set of parameter values. Under a ParameterSet, ``Run'' objects are stored, each of which has a unique random number seed.
In addition to these, we can define a post process called ``Analyzer''.
}
\end{center}
\end{figure}

OACIS stores the simulation results in a three-layered structure (``Simulator'', ``ParameterSet'', ``Run'') as shown in Fig.~\ref{fig:data-structure}.
Each simulator has several ParameterSets, and each ParameterSet has several Runs.
A simulator corresponds to an executable command to run a simulation program.
A ParameterSet represents a set of parameter values which are required by the simulator.
Let us consider a simulation model which has two input parameters $p_1$ and $p_2$, for example.
A set of values, say $ \{ p_1 = 0.2, p_2 = 0.5 \} $, corresponds to a ParameterSet.
A Run corresponds to a single Monte-Carlo run having a unique random number seed.

In addition to these three layered structures, you can define a post-process, called ``Analyzer'', which is conducted against simulation results, such as statistical analysis and visualization.
For each Simulator, we can define several Analyzers.
Likewise Simulators, Analyzers are executed at a computational host and their results are stored automatically after they finished.
A result of Analyzer is called ``Analysis''. The relation between Analyzer and Analysis is similar to that between Simulator and Run.
We can define two types of analyzers. One is analyzers conducted against the simulation result of a single Run. The other one is analyzers conducted against the simulation results for all the runs in a single ParameterSet. An Analysis are created under a Run or a ParameterSet depending on the type of the Analyzer.

\subsection{Job Execution Sequence}

\begin{figure}[ht]
\begin{center}
\includegraphics[width=0.7\columnwidth]{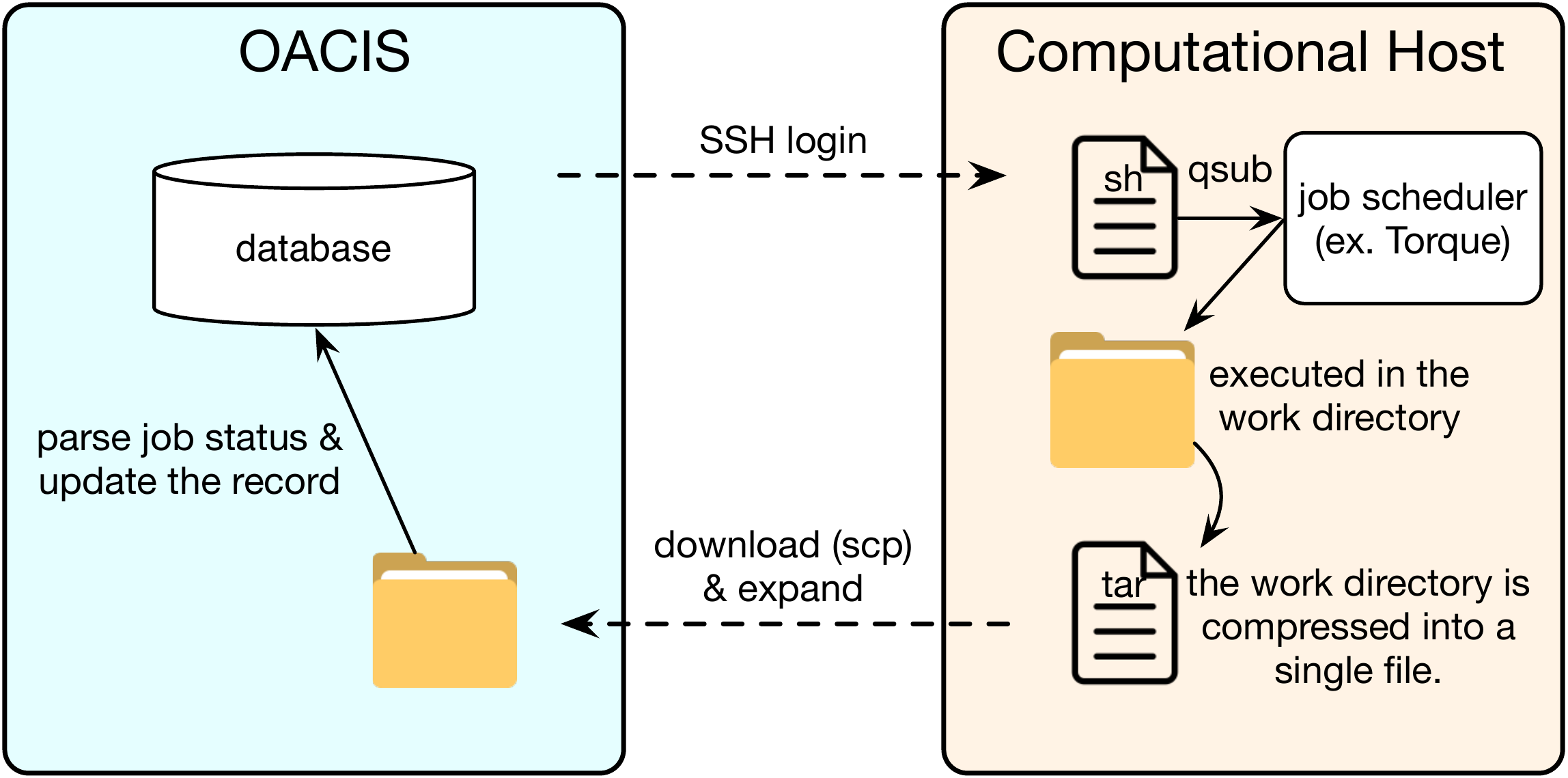}
\caption{\label{fig:job_execution_sequence}
A sequence of job executions. OACIS logs in to a computational host via SSH.
Then it makes a shell script and a temporary directory for the job.
The shell script is submitted to a job scheduler and then the job is executed in the temporary directory.
When the job finishes, the temporary directory is compressed into a single file.
OACIS downloads the compressed file and expand it to store the simulation outputs.
The logs of the job are also parsed by OACIS and recorded in the database.
}
\end{center}
\end{figure}

When you register a simulator on OACIS, you save a command line string to execute the simulation, not the execution program itself.
By this specification, OACIS can run various programs written in any programming language.
It also means that the simulation program must be compiled on computational hosts before submitting a job.
When a job is submitted, OACIS generates a shell script, which we call ``job script'', including the command line to execute the program.
For each Run, one job script is created.
Job scripts created by OACIS are submitted to the job schedulers (such as Torque) on computational hosts via SSH.
Just before submitting job scripts, a temporary directory called ``work directory'' is created for each job.
A job is executed in its work directory and the files and directories created under this directory is included into OACIS as its simulation outputs.
Thus, OACIS requires simulators to create all the output files under the current directory.
The whole sequence is illustrated in Fig.~\ref{fig:job_execution_sequence}.

When a job is executed, input parameters of the simulator are given as command line arguments or by a JSON file.
We can choose one of these when registering the simulator on OACIS.
If you choose the former one as a way to set input parameters, the parameters are given as the command line arguments, i.e., the command to execute the simulation program is concatenated by parameters and embedded into the job script.
If you choose the JSON format as a way to set input parameters, a JSON file named {\it \_input.json} is created in the work directory before execution of the jobs. Simulator must be implemented such that it reads the json file in the current directory.

\begin{figure}[ht]
\begin{center}
\includegraphics[width=0.6\columnwidth]{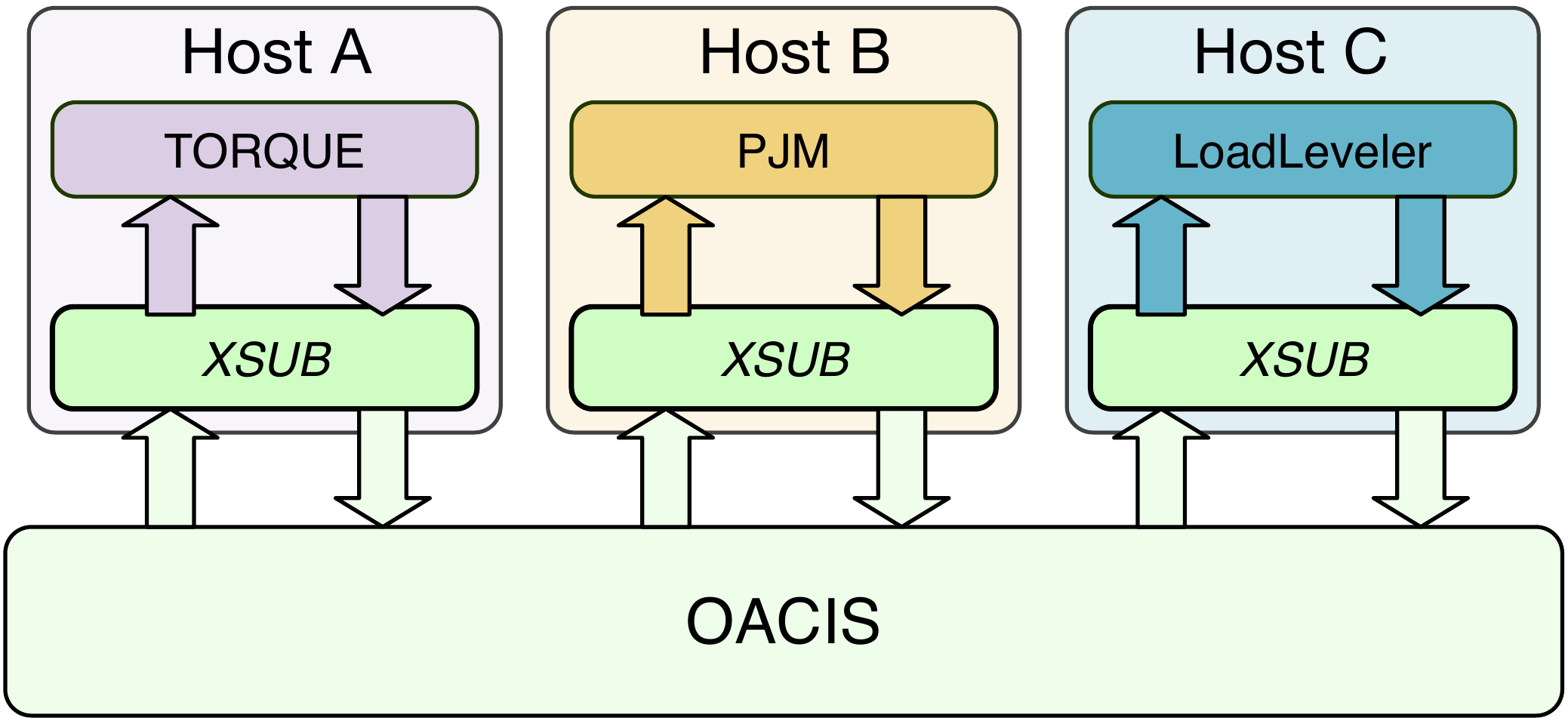}
\caption{\label{fig:xsub}
  XSUB is a wrapper script for job schedulers.
  Even though the specifications of job schedulers are different for each host, OACIS has a uniform interface against the job schedulers using XSUB.
  XSUB is designed to be extensible so that it can adapt to various types of schedulers.
}
\end{center}
\end{figure}

When using high-performance computers (HPCs), it is usually required to submit a job using a job scheduler such as Torque.
However, the specifications of the job schedulers, not only the submission command but the format of job scripts, often vary according to systems.
Job schedulers often have its own I/O format and we usually have to change job scripts to conform with its dialect.
In order to absorb the difference of the schedulers, we introduced a small script called ``XSUB'' (\verb"https://github.com/crest-cassia/xsub").
XSUB is a wrapper script to support various job schedulers and is also released as an open-source software under the MIT license.
By introducing XSUB, OACIS does not need to prepare the script specialized for each job scheduler as shown in Fig.~\ref{fig:xsub}.
A user can support new job scheduler without changing the code of OAICS, which makes the whole architecture more flexible and extensible.

\subsection{Read-Only Mode}

\begin{figure}[ht]
\begin{center}
\includegraphics[width=0.7\columnwidth]{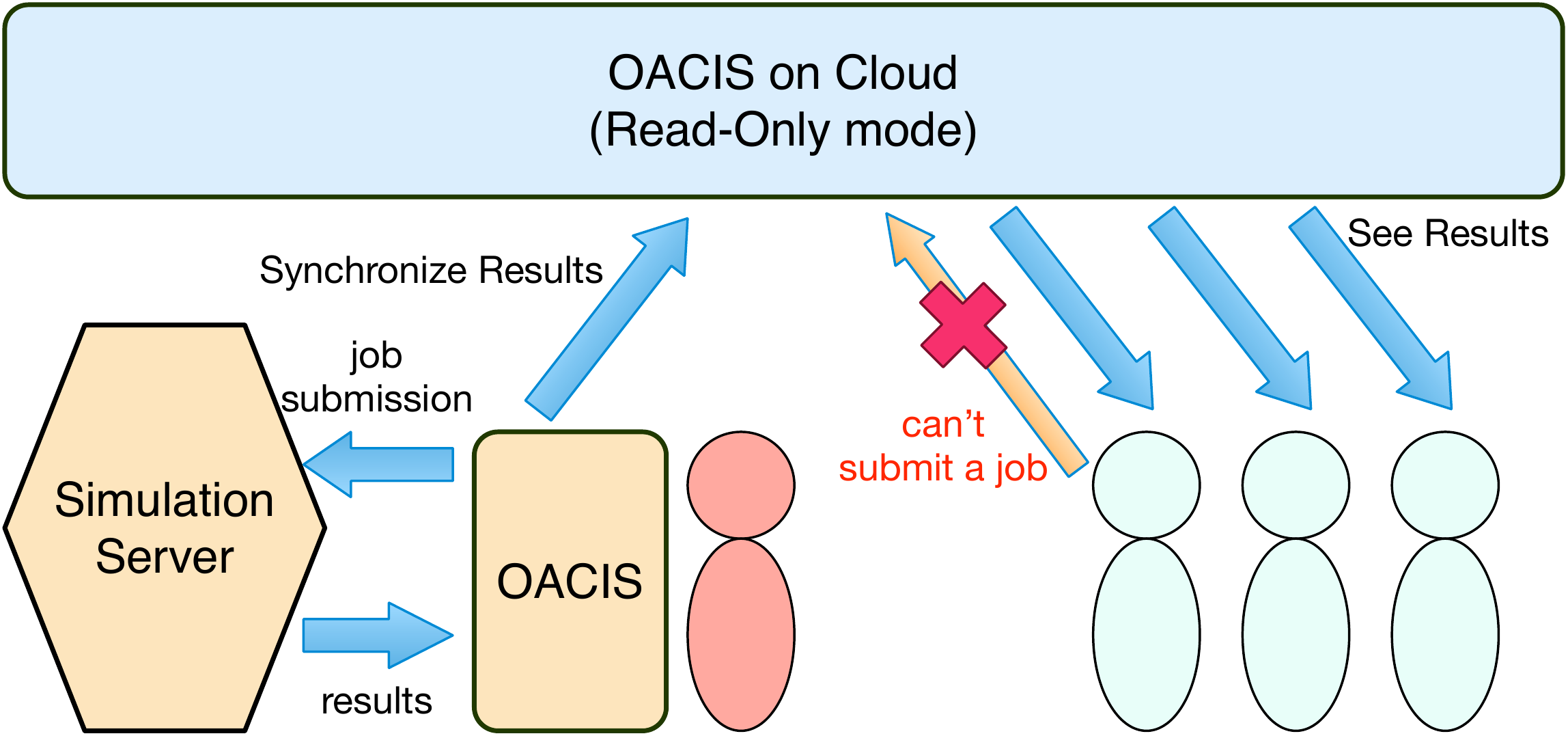}
\caption{\label{fig:readonly}
  A use case of Read-Only mode. A researcher uses OACIS to manage jobs in the intranet.
  In order to share the results with the collaborators, the researcher launched another OACIS with Read-Only mode on a cloud server, and upload the results in the local OACIS.
  By sending a URL of the public OACIS, collaborators may access to it.
}
\end{center}
\end{figure}

When you want to share your results with the researcher at a remote site, ``Read-Only'' mode of OACIS is quite useful.
If you launch OACIS in Read-Only mode, creating documents (Simulators, Parameter sets, Runs, Hosts, Analyzers, and Analyses) are prohibited, so it is not possible to submit a job or register a simulator.
Therefore, you can run an OACIS server in a public network more safely.

A typical usage of Read-Only mode would be the following (See Fig.~\ref{fig:readonly}).
A researcher executes simulations using OACIS launched at his/her local site.
After the jobs finsihed, the researcher synchronizes the data to the remote OACIS which are launched at a cloud server in Read-Only mode.
The researcher may allow the collaborators to see the results while preventing a malicious user from executing invalid commands.

\subsection{Application Programming Interfaces}

OACIS provides application programming interfaces (APIs) in Ruby and Python programming languages.
In principle, any operations on OACIS can be conducted by APIs, such as finding Simulators, creating a new ParameterSet, and deleting a Run.
Actually, the same methods are used behind the web browser interfaces.

The following is a small sample conducting a parameter sweep over parameters {\it p1} and {\it p2} of a Simulator named {\it my\_simulator}.

\definecolor{dkgreen}{rgb}{0,0.6,0}
\definecolor{gray}{rgb}{0.5,0.5,0.5}
\definecolor{mauve}{rgb}{0.58,0,0.82}
\lstset{
  frame=tb,
  language={Ruby},
  basicstyle={\small\ttfamily},
  identifierstyle={\small\ttfamily},
  keywordstyle=\color{blue},
  commentstyle=\color{dkgreen},
  stringstyle=\color{mauve}
}
\begin{lstlisting}[label=samplecode,caption=A sample script showing a usage of OACIS APIs.]
# get Simulator and Host objects
sim = Simulator.find_by_name("my_simulator")
host = Host.find_by_name("localhost")

p1_values = [1.0,2.0,3.0,4.0,5.0]
p2_values = [2.0,4.0,6.0,8.0,10.0]

# iterate over p1 and p2
p1_values.each do |p1|
  p2_values.each do |p2|
    param = {"p1"=>p1,"p2"=>p2}
    # create a ParameterSet for the given parameters
    ps = sim.find_or_create_parameter_set( param )
    # create Runs under the given ParameterSet
    runs = ps.find_or_create_runs_upto(5, submitted_to: host)
  end
end
\end{lstlisting}

Using these APIs, we can automate explorations in parameter space.
Applications include parameter sweeps, sensitivity analysis, optimization of parameters, and applying machine learning to the results of simulation results.

\section{Getting Started with OACIS}

\subsection{Installation}

To start using OACIS, we need to set up both OACIS and computational hosts.
There are two ways of setting up OACIS.
The first one is running a virtual machine image in which OACIS is already set up using Docker (\verb"https://www.docker.com/").
In the docker image, all the prerequisites are already installed.
The image is built on ``docker hub'' (\verb"https://hub.docker.com/"), a public site hosting docker images, which lets you start OACIS with a few commands.
The second one is using a native (non virtual machine) environment and setup prerequisites manually.
For the latter option, we need to use a Unix-like OS, such as Linux or macOS, and appropriate version of Ruby and MongoDB.

In this section, we briefly explain how to set up OACIS using Docker to let you try OACIS quickly.
For the full instruction on the installation, please refer to the documentation of OACIS (http://crest-cassia.github.io/oacis/en/install.html).

First, install Docker on your machine. The installation procedure depends on your platform and the official documentation of Docker tells you how to install Docker. https://docs.docker.com/engine/installation/.
After you installed Docker, run the following command to download the image and run the container. (If you are using Docker Toolbox, the command is different. Please refer to the documentation at the repository page (\verb"https://github.com/crest-cassia/oacis_docker").

\begin{verbatim}
docker run --name my_oacis -p 127.0.0.1:3000:3000 -dt oacis/oacis
\end{verbatim}

\begin{figure}[ht]
\begin{center}
\includegraphics[width=0.8\columnwidth]{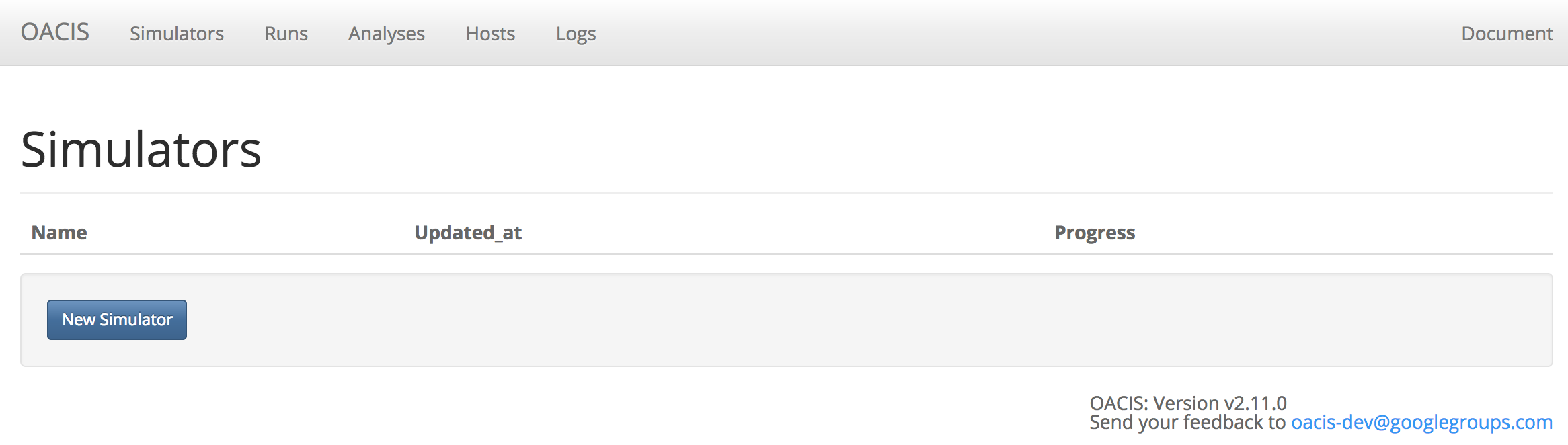}
\caption{\label{fig:top_page}
A snapshot of the top page of OACIS.
}
\end{center}
\end{figure}

When you run this command for the first time, it takes time to download the image from the public repository of docker images.
After the image is downloaded, the container (a virtual machine) is launched and OACIS is started in the container.
A few tens of seconds are necessary to complete the booting of OACIS.
Access { http://localhost:3000 } to see the top page of OACIS as in Fig.~\ref{fig:top_page}.

\subsection{Registering Hosts and Simulators}

After we installed OACIS, we need to set up computational hosts and register a simulator on OACIS.

To add a computational host, we first need to set up an SSH authentication key such that OACIS can connect to it without entering a password.
Then XSUB must be installed to submit jobs.
The detailed set up sequence are shown at a page of the OACIS documentation (\verb"http://crest-cassia.github.io/oacis").
If you are using Docker, you can skip this process since the container is already set up also as a computational host.

To submit jobs using OACIS, we need to register a simulator on OACIS.
Since we save the command line to execute the simulation, not the simulation program itself, simulators must be prepared on the computational host in advance.
In order to execute simulators from OACIS, your simulator must satisfy the following requirements.

\begin{enumerate}
  \item The output files or directories must be created in the current directory.
  \item Simulator must receive input parameters either from command line arguments or from a JSON file.
  \item The simulator must not conflict with the files {\it \_status.json, \_time.txt, \_version.txt}, which are created by OACIS in the work directory.
  \item The simulator must return 0 when finished successfully. The return code must be non-zero when an error occurs during the simulation.
\end{enumerate}

The first requirement is about the simulation outputs.
OACIS creates a temporary directory for each job and executes the job in that temporary directory.
Since all the files and directories in the temporary directory are stored in OACIS as the simulation outputs, your simulators must create output files under the current directory.
The second requirement specifies how simulators receives its input parameters given by OACIS.
The third requirements is about the temporary files. To record the information of jobs, OACIS creates several files in the work directory.
Simulators must not conflict with those. 
The fourth requirement is necessary to check if the simulator finished successfully or not.
We prepared a sample simulator in (\verb"https://github.com/yohm/sim_ns_model").

\begin{figure}[ht]
\begin{center}
  \subfloat{\includegraphics[width=0.43\columnwidth]{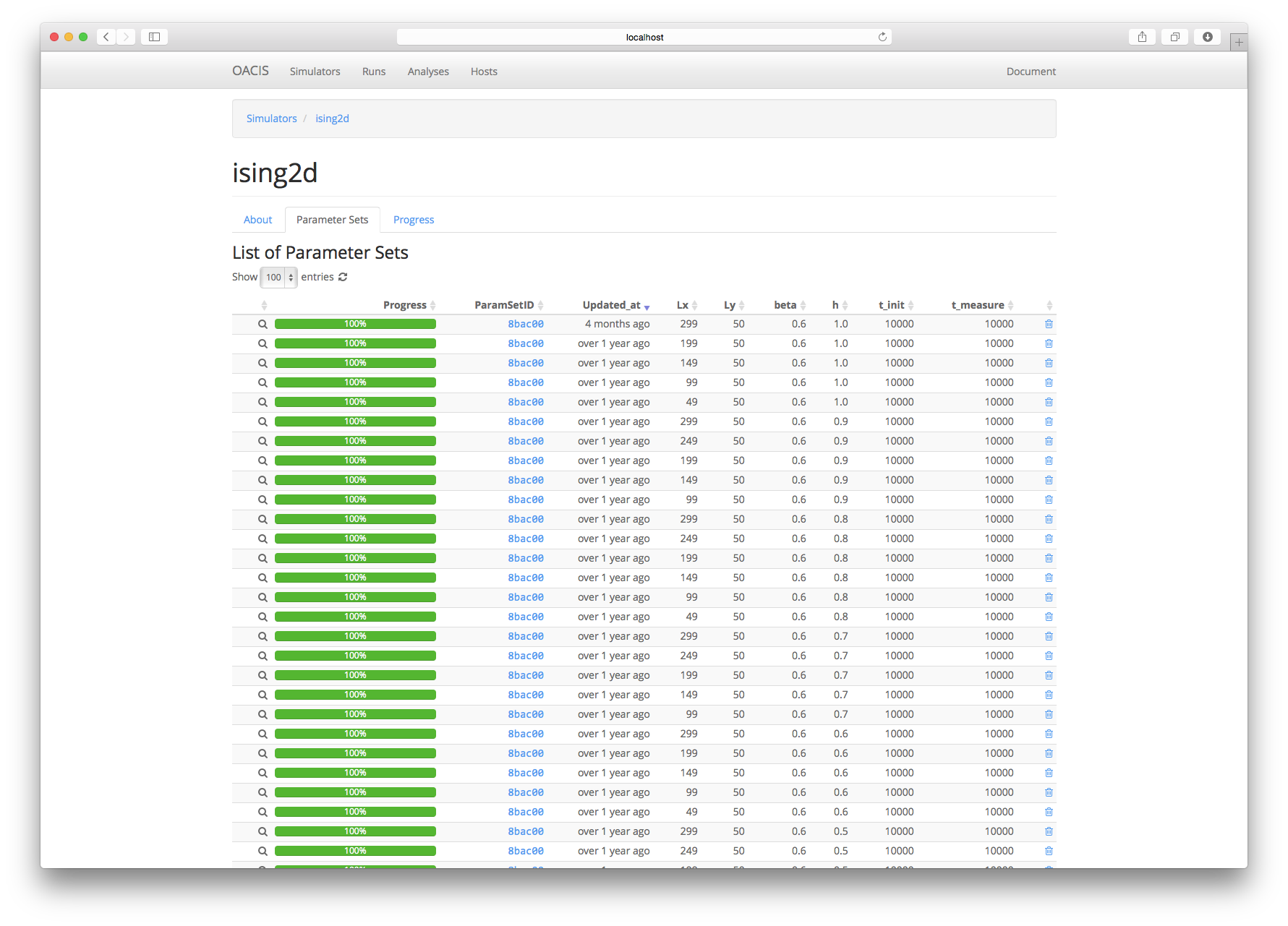}}
  \hspace{2mm}
  \subfloat{\includegraphics[width=0.43\columnwidth]{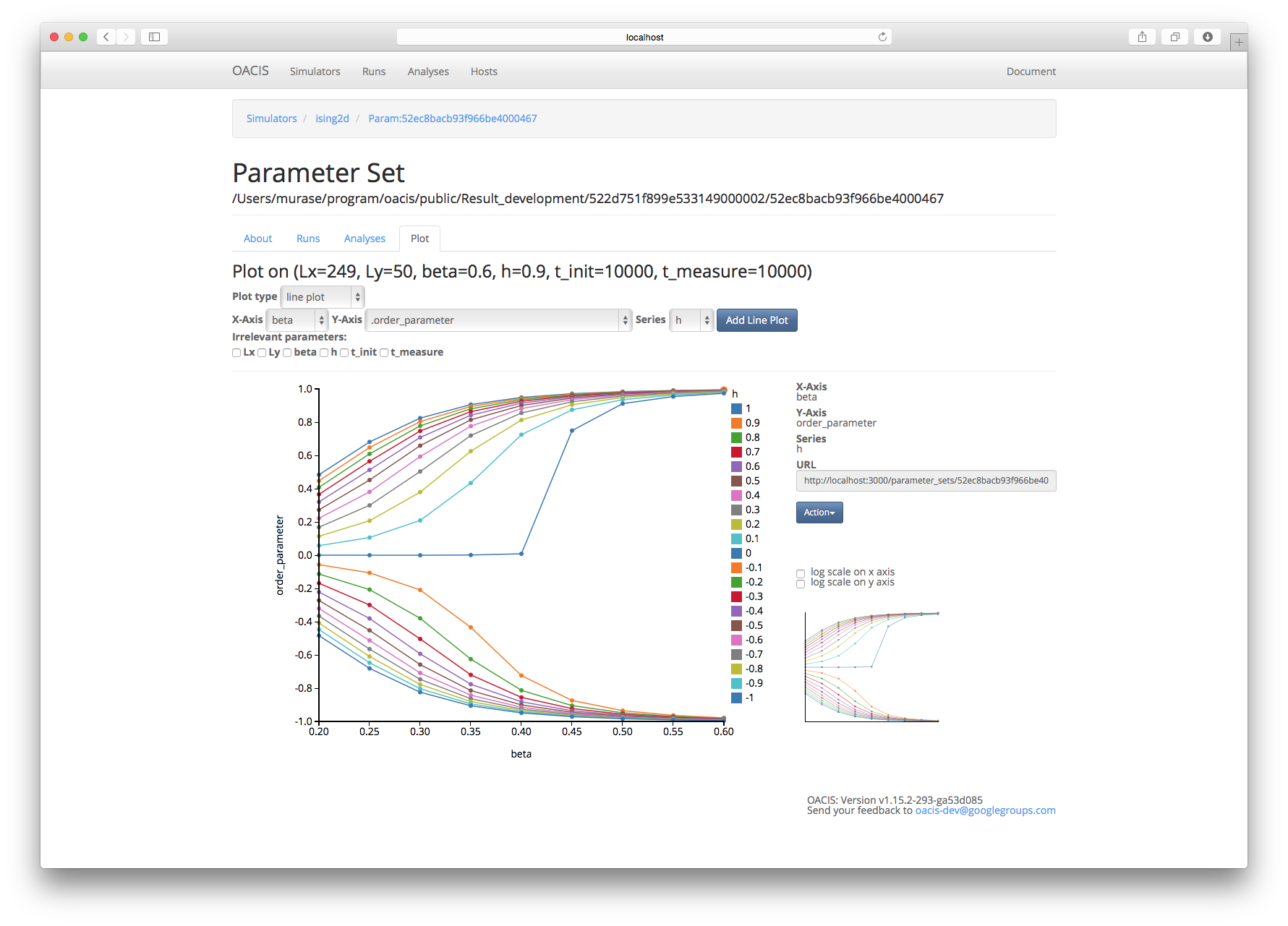}}
  \caption{\label{fig:snapshots}
  (left) A snapshot of the list of ParameterSets. (right) A snapshot of a plot made by OACIS. In this plot, a point corresponds to a ParameterSet. All the relevant ParameterSets are automatically collected by OACIS.
}
\end{center}
\end{figure}

After you prepared a simulator, you can register it on OACIS from a web browser interface.
An instruction can be found at the tutorial page of OACIS documentation (\verb"http://crest-cassia.github.io/oacis").
After simulators are registered, we can submit jobs. 
All the jobs are automatically managed by OACIS thus you can trace the simulation output even after several months.
OACIS also provides a function to make a plot, with which we are able to visually inspect how the output depends on input parameters. (See Fig.~\ref{fig:snapshots} for examples.)
The tutorial page gives you an instruction about a typical usage of OACIS using a simple sample simulator.

\section{Conclusion}

In this short paper, we presented a software framework OACIS.
It is an ongoing project so the development will continue to enhance the functionality and improve the usability even more.
If you are interested in this project, please check the latest status in our official project page (\verb"https://github.com/crest-cassia/oacis") and try to use it in your research.

\section*{acknowledgments}
This research was supported by JST CREST and by MEXT as ``Exploratory Challenges on Post-K computer(Studies of multi-level spatiotemporal simulation of socioeconomic phenomena)''.

\section*{References}
\providecommand{\newblock}{}

\end{document}